\documentclass[9pt,twocolumn,twoside]{pnas-new}
\templatetype{pnasresearcharticle} 
\setboolean{displaywatermark}{false}
\usepackage{graphicx,epsfig,chemarr}
\usepackage{amsmath,amssymb}
\usepackage{times}

\usepackage{braket}
\usepackage{booktabs}
\usepackage{eucal}
\usepackage{soul}
\def\pseudo{+}
\def\pseudo{\dashv}
\def\T{{\intercal}}
\def\be{\begin{equation}}
\def\ee{\end{equation}}
\def\ba{\begin{eqnarray}}
\def\ea{\end{eqnarray}}

\def\d{{\rm d}}

\def\eg{{\it{e.g.,~}}}
\def\ie{{\it{i.e.,~}}}

\def\lams{{\lambda_\ast}}
\def\ubs{{\ub_{\ast}}}
\def\ubsT{{\ub_{\ast}^\T}}
\def\mba{{\mb_{(\alpha)}}}

\def\mba{{\mb_{\alpha}}}

\def\Ab{{\mathbf{A}}}
\def\Bb{{\mathbf{B}}}
\def\cb{{\mathbf{c}}}
\def\Db{{\mathbf{D}}}
\def\cbs{{\cb_{\ast}}}

\def\Kb{{\mathbf{K}}}

\def\mb{{\mathbf{m}}}
\def\nb{{\mathbf{n}}}
\def\fb{{\mathbf{f}}}
\def\dfb{\mathbf{\delta{f}}}

\def\rb{{\mathbf{r}}}

\def\vb{{\mathbf{v}}}
\def\sbs{{\mathbf{s_\ast}}}
\def\ub{{\mathbf{u}}}
\def\dub{\mathbf{\delta{u}}}
\def\Zb{{\mathbf{Z}}}

\def\vbT{\vb^\T}

\def\CC{{\mathbf{C}}}
\def\UU{{\mathbf{U}}}
\def\SS{{\mathbf{S}}}
\def\Ck{{\CC_k}}
\def\Uk{{\UU_k}}
\def\Sk{{\SS_k}}
\def\MM{{\mathbf{M}}}
\def\EE{E}

\def\epij{{e_{ij}}}
\def\HH{\mathbf{H}}
\def\DH{\delta{\mathbf{H}}}
\def\DHi{\delta{\mathbf{H}_i}}
\def\DHj{\delta{\mathbf{H}_j}}
\def\DHij{\delta{\mathbf{H}_{i,j}}}
\def\DDHij{\Delta{\mathbf{H}_{i,j}}}
\def\Hs{{\HH^{\prime}}}
\def\GG{\mathbf{G}}
\def\DG{\delta{\mathbf{G}}}
\def\DGi{\delta{\mathbf{G}_i}}
\def\DGj{\delta{\mathbf{G}_j}}
\def\DGij{\delta{\mathbf{G}_{i,j}}}
\def\DDGij{\Delta{\mathbf{G}_{i,j}}}
\def\Gs{{\GG^{\prime}}}
\def\DF{\delta{F}}
\def\DFi{\delta{F_{i}}}
\def\DFj{\delta{F_{j}}}
\def\DFij{\delta{F_{i,j}}}
\def\nij{\nb_{ij}}
\def\nijT{\nij^\T}
\def\KK{\mathcal{K}}
\def\ui{{p}}
\def\uj{{q}}
\def\bi{{p'}}
\def\bj{{q'}}
\def\aH{{\rm H}}
\def\aP{{\rm P}}

\title{Green function of correlated genes in a minimal mechanical model of protein evolution}

\author[a]{Sandipan Dutta}
\author[b]{Jean-Pierre Eckmann} 
\author[c]{Albert Libchaber} 
\author[a,d,1]{Tsvi Tlusty}

\affil[a]{Center for Soft and Living Matter, Institute for Basic Science (IBS), Ulsan 44919, Korea}
\affil[b]{D\'{e}partement de Physique Th\'{e}orique and Section de
Math\'{e}matiques, Universit\'{e} de Gen\`{e}ve, CH-1211, Geneva 4,
Switzerland}
\affil[c]{Center for Studies in Physics and Biology, The Rockefeller University, 1230 York Avenue, New York, NY 10021, USA }
\affil[d]{Department of Physics, Ulsan National Institute of Science and Technology (UNIST), Ulsan 44919, Korea}

\leadauthor{Dutta}

\significancestatement{
Many protein functions involve large-scale motion of their amino acids, while alignment of their sequences shows long-range correlations. This has motivated search for physical links between these genetic and phenotypic collective behaviors. 
The major challenge is the complex nature of protein: non-random hetero-polymers made of twenty species of amino acids that fold into strongly-coupled network. In light of this complexity, simplified models that can elucidate the underlying principles are worthwhile to explore. 
Our model describes protein in terms of the Green function, which directly links the gene to force propagation and collective dynamics in the protein.  This allows for derivation of basic determinants of evolution, such as fitness landscape and epistasis, which are often hard to calculate. 
}

\correspondingauthor{\textsuperscript{1}To whom correspondence should be addressed. E-mail: tsvitlusty@gmail.com}
\keywords{Protein evolution $|$ Epistasis $|$ Genotype-to-phenotype map | Green function $|$ Dimensional reduction}

\begin{abstract}
The function of proteins arises from cooperative interactions and rearrangements of their amino acids, which exhibit large-scale dynamical modes. 
Long-range correlations have also been revealed in protein sequences, and this has motivated the search for physical links between the observed genetic and dynamic cooperativity. 
We outline here a simplified theory of protein, which relates sequence correlations to physical interactions and to the emergence of mechanical function.  
Our protein is modeled as a strongly-coupled amino acid network whose interactions and  motions are captured by the mechanical propagator, the Green function. 
The propagator describes how the gene determines the connectivity of the amino acids, and thereby the transmission of forces. 
Mutations introduce localized perturbations to the propagator which scatter the force field. 
The emergence of function is manifested by a topological transition when a band of such perturbations divides the protein into subdomains. 
We find that epistasis -- the interaction among mutations in the gene -- is related to the nonlinearity of the Green function, which can be interpreted as a sum over multiple scattering paths.  
We apply this mechanical framework to simulations of protein evolution, and observe long-range epistasis which facilitates collective functional modes. 

\end{abstract}
\dates{\today}
\doi{Green function of correlated genes in a minimal mechanical model of protein evolution}

\begin{document}
\verticaladjustment{-2pt}
\maketitle
\thispagestyle{firststyle}
\ifthenelse{\boolean{shortarticle}}{\ifthenelse{\boolean{singlecolumn}}{\abscontentformatted}{\abscontent}}{}

\begin{figure*}[p]
\centering
\includegraphics[width=1.0\textwidth]{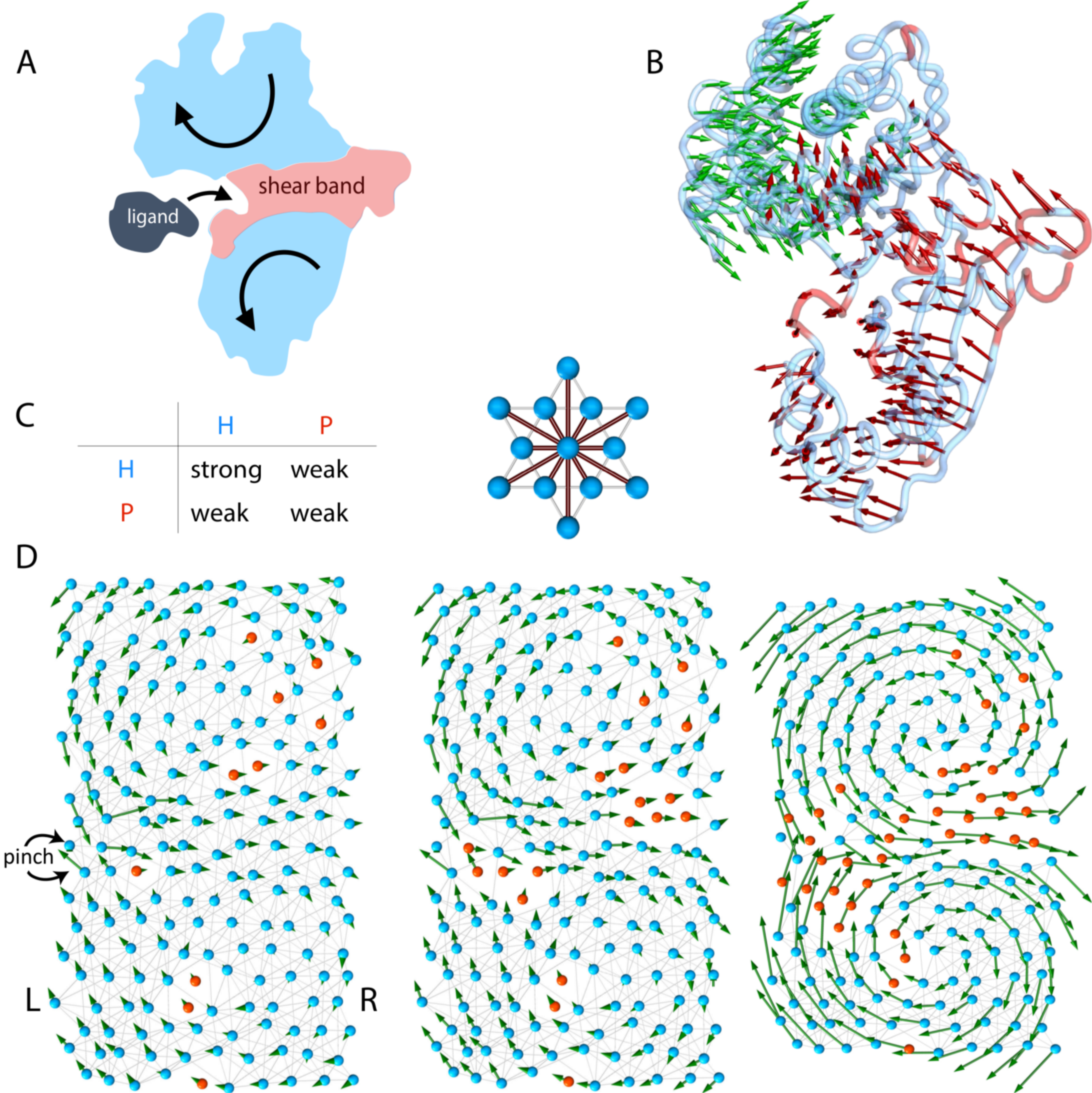}
\caption{\textbf{Protein as an evolving machine and propagation of mechanical forces.} 
(A) Formation of a softer `shear band' (red) separating the protein into two rigid subdomains (light blue). When a ligand binds, the biochemical function involves a low-energy, hinge-like or shear motion (arrows).  
(B) Shear band and large-scale motion in a real protein:  the arrows show the displacement of all amino acids (AAs) in  human glucokinase when it binds glucose (PDB structures: 1v4s and 1v4t). The coloring shows a high shear region (red) separating two  low-shear domains that move as rigid bodies (shear calculated as in \cite{Mitchell2016,Tlusty2017}).
\textbf{The mechanical model.}  (C) The protein is made of two species of AAs, polar ($\aP$, red) and hydrophobic ($\aH$, blue) whose sequence is encoded in a gene. Each AA forms weak or strong bonds with its $12$ near neighbors (right) according to the interaction rule in the table (left).   
(D) The protein is made of   $10 \times 20 = 200$ AAs whose positions are randomized from a regular triangular lattice. Strong bonds are shown as gray lines. Evolution begins from a random configuration (left) and evolves by mutating one AA at each step, switching between $\aH$ and $\aP$. The fitness is the mechanical response to a localized force probe (`pinch') [\ref{eq:fitness}]. After ${\sim}10^3$ mutations (middle: intermediate stage) the evolution reaches a solution (right). The green arrows show the mechanical response: a hinge-like, low-energy motion with a shear band starting at the probe and traversing the protein, qualitatively similar to (B). 
}
\label{fig:1}
\end{figure*}

\dropcap{A} common physical basis for the diverse biological functions of proteins is the emergence of collective patterns of forces and coordinated displacements of their amino acids (AAs) \cite{Daniel2003,Bustamante2004,Hammes-Schiffer2006,Boehr2006,Bahar2010a,Karplus2002,Henzler-Wildman2007,Huse2002,Eisenmesser2005,Savir2007,Goodey2008,Savir2010a,Grant2010}.  
In particular, the mechanisms of allostery \cite{Monod1965,Perutz1970,Cui2008,Daily2008,Motlagh2014} and induced fit \cite{Koshland1958} often involve global conformational changes by hinge-like rotations, twists or shear-like sliding of protein subdomains \cite{Gerstein1994,Mitchell2016,Mitchell2017}.
An approach to examine the link between function and motion is to model proteins as elastic networks 
\cite{Tirion1996,Chennubhotla2005,Bahar2010,Lopez-Blanco2016}. Decomposing the dynamics of the network into normal modes revealed that  low-frequency  `soft' modes capture functionally relevant large-scale motion \cite{Levitt1985,Tama2001,Bahar2005,Haliloglu2015}, especially in allosteric proteins \cite{Ming2005,Zheng2006,Arora2007}.
Recent works associate these soft modes with the emergence of  weakly connected regions in the protein (Fig.~\ref{fig:1}A,B) --  `cracks', `shear bands'  or `channels'  \cite{Mitchell2016,Mitchell2017,Miyashita2003,Tlusty2016,Tlusty2017} -- that enable viscoelastic motion \cite{Qu2013,Joseph2014}.  
Such `floppy' modes \cite{Alexander1998,Alexander1982,Phillips1985,Thorpe2001}  emerge  in models of allosteric proteins \cite{Hemery2015,Flechsig2017,Tlusty2017,Thirumalai2017} and networks \cite{Rocks2017,Yan2017,Yan2017a}.

Like their dynamic phenotypes, proteins' genotypes are remarkably collective.  When aligned, sequences of protein families show long-range correlations among the AAs \cite{Goebel1994,Marks2011,Marks2012,Jones2012,Lockless1999,Suel2003,Reynolds2011,Hopf2017,Poelwijk2017,Halabi2009,Rivoire2016,Tesileanu2015,Juan2013}.
The correlations indicate \emph{epistasis}, the interaction among mutations that take place among residues linked by physical forces or common function. By inducing non-linear effects, epistasis shapes the protein's fitness landscape \cite{Cordell2002,Phillips2008,Mackay2014,Ortlund2007,Breen2012,Gong2013,Miton2016}. 
Provided with sufficiently large data, analysis of sequence variation can predict the 3D structure of proteins \cite{Marks2011,Marks2012,Jones2012}, allosteric pathways \cite{Lockless1999,Suel2003,Reynolds2011}, epistatic interactions \cite{Hopf2017,Poelwijk2017} and coevolving subsets of AAs \cite{Fitch1970, Halabi2009,Rivoire2016,Tesileanu2015}. 

Still, the mapping between sequence correlation and collective dynamics -- and in particular the underlying epistatis -- are not fully understood. 
Experiments and simulations provide valuable information on protein dynamics, and extensive sequencing accumulates databases required for reliable analysis, but there remain inherent challenges: the complexity of the physical interactions and the sparsity of the data. The genotype-to-phenotype map of proteins connects spaces of huge dimension, which are hard to sample, even by high throughput experiments or natural evolution \cite{Koonin2002,Povolotskaya2010,Liberles2012}.
A complementary approach is the application of simplified coarse-grained models, such as lattice proteins \cite{Dill1985,Lau1989,Shakhnovich1991} or elastic networks \cite{Chennubhotla2005}, which allow one to extensively survey of the map and examine basic questions of protein evolution. 
We have recently used coarse-grained models to study the evolution of allostery in proteins and the geometry of the genotype-to-phenotype map \cite{Tlusty2016,Tlusty2017}.
Our aim here is orthogonal: to construct a simplified model of how the collective dynamics of functional proteins directs their evolution, and in particular to give a mechanical interpretation of epistasis.

The present paper introduces a coarse-grained theory that treats protein as an evolving amino acid network whose topology is encoded in the gene. Mutations that substitute one AA by another tweak the interactions, allowing the network to evolve towards a specific mechanical function: in response to a localized force, the protein will undergo a large-scale conformational change (Fig.~\ref{fig:1}C,D). 
We show that the application of a Green function \cite{Green1828,Abrikosov1963}  is a natural way to understand the protein's collective dynamics. 
The Green function measures how the protein responds  to a localized impulse via propagation of forces and motion. 
The propagation of mechanical response across the protein defines its fitness and directs the evolutionary search. 
Thus, the Green function explicitly defines the map: ${\rm gene}\to{\rm amino~acid~network}\to{\rm protein~dynamics}\to{\rm function}$. 
We use this map to examine the effects of mutations and epistasis.
A mutation perturbs the Green function, and scatters the propagation of force through the protein (Fig. \ref{fig:2}). We quantify epistasis in terms of  "multiple scattering" pathways. 
These indirect physical interactions appear as long-range correlation in the coevolving genes.

Using a Metropolis-type evolution algorithm, solutions are quickly found, typically after ${\sim} 10^3$ steps. Mutations add localized perturbations to the AA network, which are eventually arranged by evolution into a continuous shear band. 
Protein function is signaled by a topological transition which occurs when a shearable band of weakly connected AAs separates the protein into rigid subdomains. 
The set of solutions is sparse: there is a huge reduction of dimension between the space of genes to the spaces of force and displacement fields. 
We find a tight correspondence between correlations in the genotype and phenotype.  Owing to its mechanical origin, epistasis becomes long ranged along the high shear region of the channel.

\section*{Model: Protein as an evolving machine}
\subsection*{The amino acid network and its Green function}
We use a coarse grained description in terms of an elastic network \cite{Tirion1996,Chennubhotla2005,Bahar2010,Lopez-Blanco2016,Levitt1985,Alexander1998} whose connectivity and interactions are encoded in a gene (Fig.~\ref{fig:1}C,D). 
Similar vector elasticity models, discrete  and continuous,  were considered in \cite{Tlusty2016} and \cite{Tlusty2017} (see app. B3 therein).
The protein is a chain of $n_a=200$ amino acids, $a_i$ ($i=1,\dotsc ,n_a$) folded into a $10 \times 20$  two-dimensional hexagonal lattice ($d=2$). We follow the HP model \cite{Lau1989,Dill1985} with its two species of AAs, hydrophobic ($a_i = \aH$) and polar ($a_i=\aP$). The AA chain is encoded in a gene $\cb$, a sequence of $200$ binary codons, where $c_i=1$  encodes a $\aH$ AA and   $c_i=0$ encodes a $\aP$.  

We consider a constant fold, so any particular codon $c_i$ in the gene encodes an AA $a_i$ at a certain constant position $\rb_i$   in the protein.  The positions $\rb_i$ are randomized to make the network amorphous. These $n_d =d \cdot n_a=400$ degrees-of-freedom are stored in a vector $\rb$.  Except the ones at the boundaries, every AA is connected by harmonic springs to $z=12$ nearest and next-nearest neighbors. There are two flavors of bonds according to the chemical interaction which is defined as an ${\rm AND}$ gate: a strong $\aH {-}\aH$ bond and weak $\aH {-}\aP$ and $\aP {-}\aP$ bonds. 
The strength of the bonds determines the mechanical response of the network to a displacement  field $\ub$, when the AAs are displaced as $\rb_i \to \rb_i + \ub_i$. The response is  captured by Hooke's law that gives the force field $\fb$ induced by a displacement field,
$\fb =\HH(\cb) \, \ub \,.$
The analogue of the spring constant is the Hamiltonian  $\HH(\cb)$, a $n_d {\times} n_d$ matrix, 
which records the connectivity of the network and the strength of the bonds. $\HH(\cb)$ is a nonlinear function of the gene $\cb$, reflecting the AA interaction rules of Fig. \ref{fig:1}C  (see [\ref{eq:Hamiltonian}], Methods).

\begin{figure}[ht]
\centering
\includegraphics[width=1.0\linewidth]{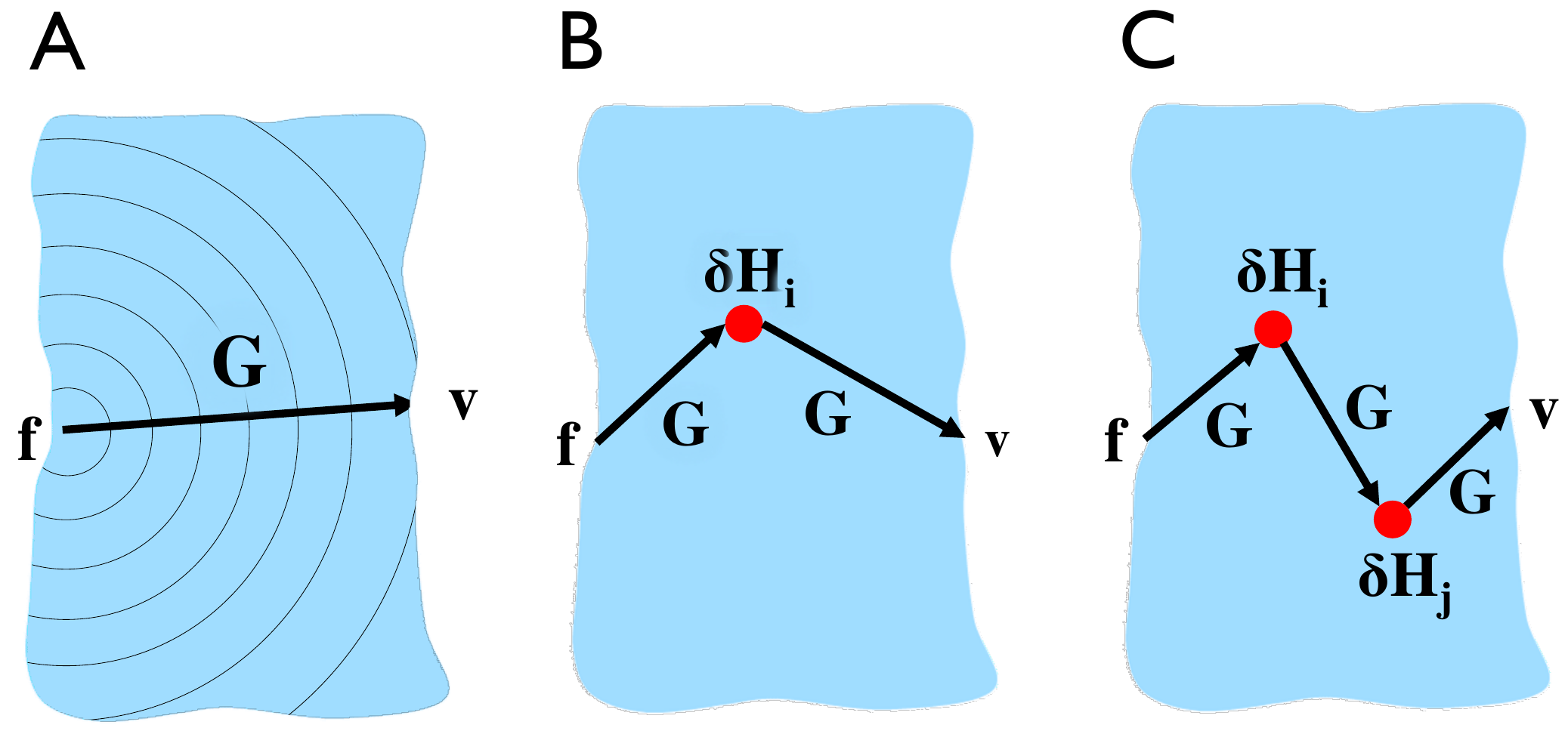}
\caption{\textbf{Force propagation, mutations and epistasis.}
(A) The Green function $\GG$ measures the propagation the mechanical signal,  depicted as a ``diffraction wave", across the protein (blue) from the force source $\fb$ (pinch) to the response site $\vb$.
(B) A mutation  $\DHi$ deflects the propagation of force. The effect of the mutation on the propagator $\DG$ can be described as a series of multiple scattering paths [\ref{eq:Dyson}]. 
(C) The epistasis between two mutations, $\DHi$ and $\DHj$, is equivalent to a series of multiple scattering paths [\ref{eq:multiple}]. }
\label{fig:2}
\end{figure}

Evolution searches for a protein which will respond  by  a prescribed large-scale motion  to a given localized force  $\fb$ (`pinch').
In induced fit, for example, specific binding of a substrate should induce global deformation of an enzyme.
The response $\ub$ is determined by the Green function $\GG$ \cite{Green1828}, 
\begin{equation} \label{eq:Green}
\ub = \GG(\cb) \, \fb \,.
\end{equation}$\GG$ is the mechanical propagator that  measures the transmission of signals from the force source $\fb$ across the protein (Fig. \ref{fig:2}A).
Equation [\ref{eq:Green}] constitutes an explicit \emph{genotype-to-phenotype map} from the genotype $\cb$ to the mechanical phenotype $\ub$:  $\cb \to \ub(\cb) = \GG(\cb) \fb$. This reflects the dual nature of the Green function $\GG$: in the phenotype space, it is the linear mechanical propagator which turns a force into motion, $\ub = \GG \,\fb$, whereas it is also the nonlinear function which maps the gene into a propagator, $\cb\to \GG(\cb)$. 

When the protein is moved  as a rigid body, the lengths of the bonds do not change and the elastic energy cost vanishes.  A 2D protein has $n_0 =3$ such zero modes (Galilean symmetries), two translations and one rotation, and  $\HH$  is therefore always singular.
Hence, Hooke's law and [\ref{eq:Green}] imply that $\GG$ is the pseudo-inverse of the Hamiltonian, $\GG(\cb)=\HH(\cb)^\pseudo$ \cite{Penrose1955,Ben-Israel2003}, which amounts to inversion of $\HH$ in the non-singular sub-space of the  $n_d-n_0=397$ non-zero modes (Methods). 
A related quantity is the resolvent, $\GG(\omega) = ( \omega-\HH)^{-1}$, whose poles are at the energy levels of $\HH$, $\omega=\lambda_k$. 

 \begin{figure*}[ht]
\centering
\includegraphics[width=0.93\linewidth]{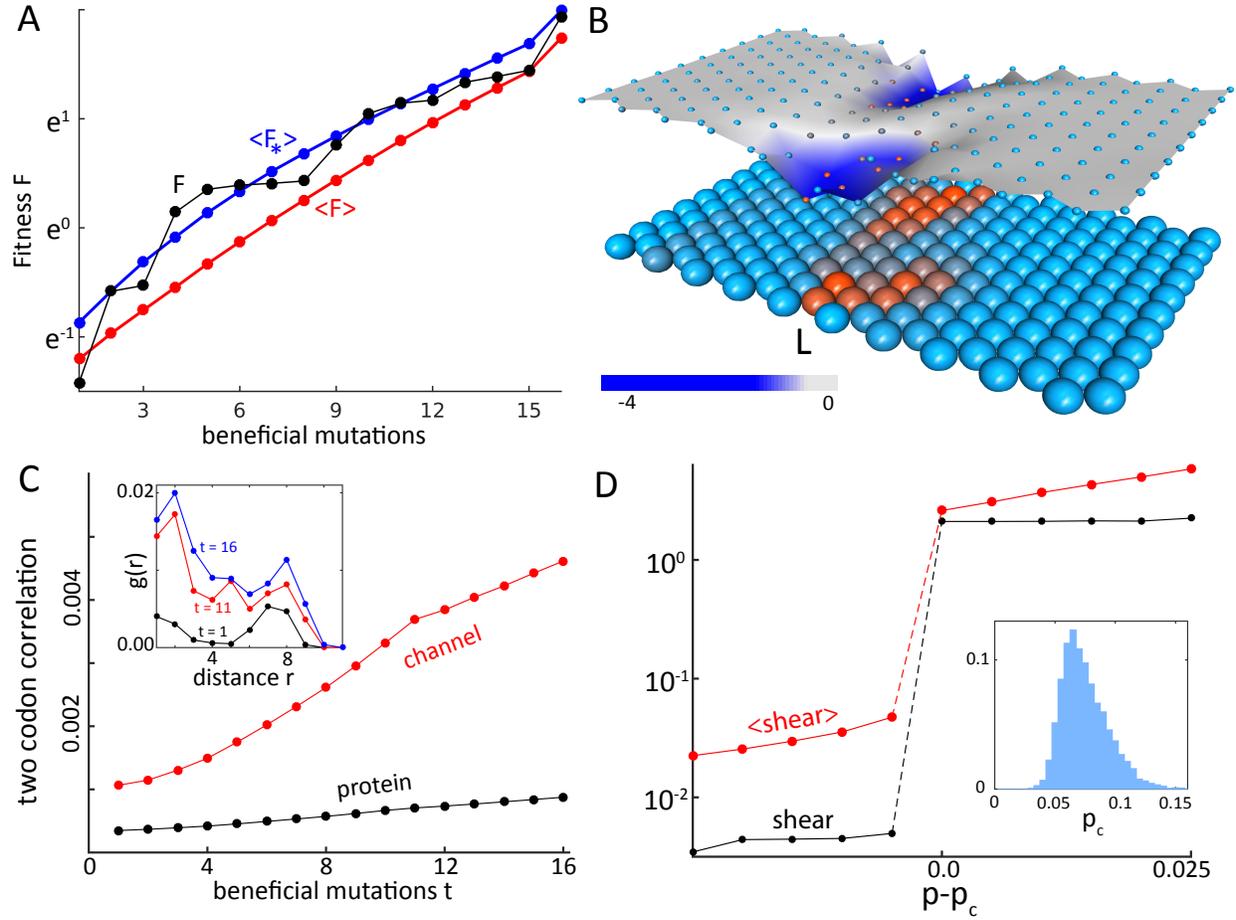}
\caption{\textbf{The mechanical  Green function and the emergence of protein function.}
(A) Progression of the fitness $F$ during the evolution run shown in Fig.~\ref{fig:1}D (black), together with the fitness trajectory averaged over ${\sim} 10^6$ runs $\langle F \rangle$ (red). Shown are the last 16 beneficial mutations towards the formation of the channel. The contribution of the emergent low-energy mode $\langle F_\ast \rangle$ (blue) dominates the fitness [\ref{eq:divergence}]. 
(B) Landscape of the fitness change $\DF$ [\ref{eq:deltaF}], averaged over ${\sim} 10^6$ solutions, for all 200 possible positions of point mutations at a solution.  Underneath, the average AA configuration of the protein is shown in shades of red ($\aP$) and blue ($\aH$).
In most sites, mutations are neutral, while mutations in the channel are deleterious on average.
(C) The average magnitude of the two-codon correlation $|Q_{ij}|$ [\ref{eq:correlation}] in the shear band (AAs in rows $7{-}13$, red) and in the whole protein (black) as a function of the number of beneficial mutations, $t$.
Inset: profile of the spatial correlation $g(r)$ within the shear band (after $t=1,11,16$ beneficial mutations).  
(D)  The mean shear in the protein in a single run (black) and averaged over ${\sim}10^6$ solutions (red), as a function of the fraction of $\aP$ amino acids, $p$. The values of $p$ are shifted by the position of the jump, $p_c$. 
  Inset: distribution of $p_c$.}
\label{fig:3}
\end{figure*}

The fitness function rewards strong mechanical response to a localized probe (`pinch', Fig.~\ref{fig:1}D):
a force dipole at two neighboring AAs, $\bi$ and $\bj$ on the left side of the protein (L), $\fb_\bj=-\fb_\bi$.
 The prescribed motion is specified by a displacement vector $\vb$, with a dipolar response,  $\vb_\uj=-\vb_\ui$, on the right side of the protein (R).
The protein is fitter if the pinch $\fb$ produces a large deformation in the direction specified by $\vb$.
To this end, we evolve the AA network to increase a
fitness function $F$, which is the projection of the displacement  $\ub = \GG \fb$ on the prescribed response $\vb$,
\begin{equation} \label{eq:fitness}
F(\cb)=\vbT\ub  =\vbT \GG(\cb) \, \fb \,.
\end{equation}  
Equation [\ref{eq:fitness}] defines the \emph{fitness landscape} $F(\cb)$.
Here we examine particular examples for a localized `pinch' $\fb$ and prescribed response $\vb$, which drive the emergence of a hinge-like mode. 
The present approach is general and can as well treat more complex patterns of force and motion. 
 
 \subsection*{Evolution searches in the mechanical fitness landscape}
Our simulations search for a prescribed response $\vb$ induced by a force $\fb$ applied at a specific site on the L side (pinch). The prescribed dipolar response may occur at any of the sites on the R side. This gives rise to a wider shear band that allows the protein to perform general mechanical tasks \footnote{Unlike a specific allostery task: communicating between two specified sites on L and R.}.
We  define the fitness as the maximum of $F$ [\ref{eq:fitness}] over all potential locations of the channel's output (typically $8{-}10$ sites, Methods). 
The protein is evolved via a point mutation process where, at each step, we flip a randomly selected codon between $0$ and $1$. This corresponds to exchanging  $\aH$ and $\aP$ at a random position in the protein, thereby changing the bond pattern and the elastic response by softening or stiffening the AA network.

Evolution starts from a random protein configuration, encoded in a random gene. Typically, we take a small fraction of AA of type $\aP$ (about $5\%$), randomly positioned within a majority of $\aH$. 
(Fig.~\ref{fig:1}D, Left). The high fraction of strong bonds renders the protein stiff, and therefore of low initial fitness $F\simeq 0$. At each step, we calculate the change in the Green function $\DG$ (by a method explained below) and use it to evaluate from [\ref{eq:fitness}] the fitness change $\DF$,
\begin{equation} \label{eq:deltaF}
\DF = \vbT \DG\, \fb~.
\end{equation}
The  fitness change $\DF$ determines the fate of the mutation: we accept the mutation if $\DF \ge
0$, otherwise the mutation is rejected.  Since fitness is measured by the criterion of strong mechanical response, it induces softening of the AA network. 

The typical evolution trajectory lasts about $10^3$ steps. Most are neutral mutations ($\DF \simeq 0$) and deleterious ones ($\DF<0$); the latter are rejected. About a dozen or so beneficial mutations ($\DF > 0$) drive the protein towards the solution (Fig.~\ref{fig:3}A). 
The increase in the fitness reflects the gradual formation of the channel, while the jump in the shear signals the emergence of the soft mode.
The first few beneficial mutations tend to form weakly-bonded, $\aP$-enriched regions near the pinch site on the L side, and close to the R boundary of the protein. The following ones join these regions into a floppy channel (a shear band) which  traverses the protein from L to R.  
We stop the simulation when the fitness reaches a large positive value $F_{\rm m} \sim 5$. The corresponding gene $\cbs$ encodes the functional protein. The ad-hoc value $F_{\rm m}\sim 5$ signals slowing down of the fitness curve towards saturation at  $F>F_{\rm m} $, as the channel has formed and now only continues to slightly widen. In this regime, even a tiny pinch will easily excite a large-scale motion with a distinct high-shear band (Fig.~\ref{fig:1}D right).

\section*{Results}
\begin{figure*}[t]
\centering
\includegraphics[width=0.85\textwidth]{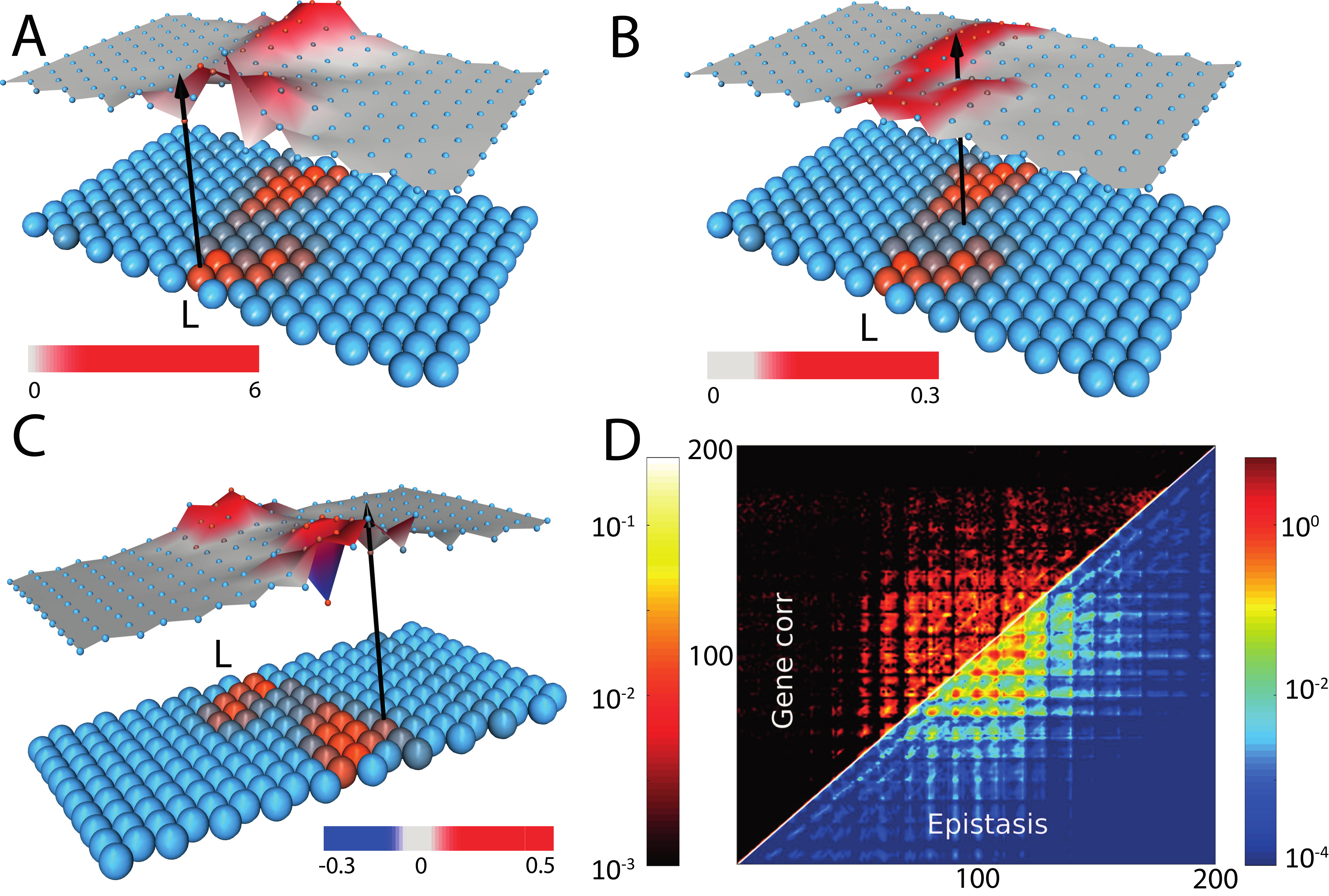}
\caption{\textbf{Mechanical Epistasis.}
The epistasis [\ref{eq:epistasis}], averaged over $\sim 10^6$ solutions $\EE_{ij}=\langle \epij \rangle$, between a fixed AA at position $i$ (black arrow) and all other positions $j$. Here, $i$ is located at (A) the binding site, (B) the center of the channel, and (C) slightly off the channel. Underneath, the average AA configuration of the protein is drawn in shades of red ($\aP$) and blue ($\aH$).
Significant epistasis mostly occurs along the $\aP$-rich channel, where mechanical interactions are long ranged. 
Though epistasis is predominantly positive, negative values also occur, mostly at the boundary of the channel (C). Landscapes are plotted for specific output site at R.
(D) The two-codon correlation function $Q_{ij}$ [\ref{eq:correlation}] measures the coupling between mutations at positions $i$ and $j$ [\ref{eq:correlation}].  The epistasis $\EE_{ij}$ and the gene correlation $Q_{ij}$ show similar patterns. Axes are the positions of $i$ and $j$ loci.  Significant correlations and epistasis occur mostly in and around the channel region (positions ${\sim}70{-}130$, rows $7{-}13$).}
\label{fig:4}
\end{figure*}

\subsection*{Mechanical function emerges at a topological transition}

The hallmark of evolution reaching a solution gene $\cbs$ is the emergence of a new zero energy-mode, $\ubs$,  in addition to the three Galilean symmetry modes.  Near the solution, the energy of this mode $\lams$ almost closes the spectral gap, $\lams \to 0$, and $\GG(\omega)$ has a pole at $\omega \approx 0$.  As a result, the emergent mode dominates the Green function, $\GG \simeq \ubs \ubsT/\lams$.
The response to a pinch will be mostly through this soft mode, and the fitness [\ref{eq:fitness}] will diverge as 
\begin{equation}
\label{eq:divergence}
F(\cbs) \simeq F_\ast= \frac{\left( \vbT \ubs\right)\left( \ubsT \fb  \right)}{\lams}~. 
\end{equation}

On average, we find that the fitness increases exponentially with the number of beneficial mutations (Fig.~\ref{fig:3}A). However, beneficial mutations are rare, and are separated by long stretches of neutral mutations. This is evident from the fitness landscape 
(Fig.~\ref{fig:3}B), which shows that in most sites the effect of mutations is practically neutral.
  The vanishing of the spectral gap, $\lams \to 0$, manifests as a topological change in the system: the AA network is now divided into two domains that can move independently of each other at low energetic cost. 
 The soft mode appears at a \emph{dynamical phase transition}, where the average shear in the protein jumps abruptly as the channel is formed and the protein can easily deform in response to the force probe (Fig. \ref{fig:3}D).

As the shear band is taking shape, the correlation among codons builds up. To see this, we align genes from the  ${\sim} 10^6$ simulations, in analogy to sequence alignment of real protein families \cite{Goebel1994,Marks2011,Marks2012,Jones2012,Lockless1999,Suel2003,Reynolds2011,Hopf2017,Poelwijk2017,Halabi2009,Rivoire2016,Tesileanu2015,Juan2013},
albeit without the phylogenetic correlation which hampers the analysis of real sequences.
At each time step we calculate the two-codon correlation $Q_{ij}$ between all pairs of codons $c_i$ and $c_j$ , 
\begin{equation}
\label{eq:correlation}
Q_{ij} \equiv \langle c_i c_j\rangle-\langle c_i \rangle \langle c_j\rangle~,
\end{equation}
where brackets denote ensemble averages. We find that most of the correlation is concentrated in the forming channel (Fig. \ref{fig:3}C), where it is tenfold larger than in the whole protein. In the channel, there is significant long-range correlation shown in the spatial profile of the correlation $g(r)$ (Fig. \ref{fig:3}C, inset).
Analogous regions of coevolving residues appear in real protein families \cite{Lockless1999,Suel2003,Reynolds2011,Halabi2009,Rivoire2016,Tesileanu2015}, as well as in 
models of protein allostery \cite{Hemery2015,Flechsig2017,Tlusty2016,Tlusty2017} and allosteric  networks \cite{Rocks2017,Yan2017}.

\subsection*{Point mutations are localized mechanical perturbations}
A mutation may vary the strength of no more than $z=12$ bonds around the mutated AA (Fig. \ref{fig:2}B).
The corresponding perturbation of the  Hamiltonian $\DH$ is therefore localized, akin to a defect  in a crystal \cite{Tewary1973,Elliott1974}. 
The mechanics of mutations can be further explored  by examining the perturbed Green function, $\Gs = \GG +\DG$,  which obeys the Dyson equation \cite{Dyson1949,Abrikosov1963} (Methods),
\begin{equation} 
\label{eq:Dyson}
\Gs = \GG - \GG \,\DH \,\Gs~.
\end{equation}
The latter can be iterated into an infinite series,
\begin{equation*}
\DG =\Gs -\GG=-\GG \,\DH \,\GG + \GG \,\DH \,\GG\,\DH \,\GG -\dotsb~.
 \end{equation*}
This series has a straightforward physical interpretation as a sum over multiple scatterings:
As a result of the mutation,  the elastic force field is no longer balanced by the imposed force $\fb$, 
leaving a residual force field $\dfb = \DH\, \ub = \DH \, \GG\, \fb$.
The first scattering  term in the series balances $\dfb$  by the deformation $\dub = \GG \,\dfb = \GG\, \DH \,\GG \fb$. Similarly, the second scattering term accounts for  further deformation induced by $\dub$, and so forth.
In practice, we calculate the mutated Green function using the Woodbury formula [\ref{eq:Woodbury}], which exploits the localized nature of the perturbation to accelerate the computation by a factor of ${\sim}10^4$(Methods).

\subsection*{Epistasis links protein mechanics to genetic correlations}
Our model provides a calculable definition of epistasis, the nonlinearity of the fitness effect of interacting mutations  (Fig \ref{fig:2}C).
We take a functional protein obtained from the evolution algorithm and mutate an AA at a site $i$. This mutation induces a change in the Green function  $\DGi$ (calculated by [\ref{eq:Woodbury}]) and hence in the fitness function $\DFi$ [\ref{eq:deltaF}].
One can similarly perform another independent mutation at a site $j$, producing a second deviation, $\DGj$ and $\DFj$. Finally, starting again from the original solution one mutates both $i$ and $j$ simultaneously, with a combined effect $\DGij$ and $\DFij$. The epistasis $\epij$ measures the departure of the double
mutation from additivity of two single mutations, 
\begin{equation}
\label{eq:epistasis}
\epij \equiv \DFij - \DFi - \DFj~.
\end{equation}
To evaluate the average epistatic interaction among AAs, we perform the double mutation calculation for all $ 10^6$ solutions and take the ensemble average $\EE_{ij}=\langle \epij \rangle$.  Landscapes of $\EE_{ij}$ show significant epistasis in the channel (Fig.~\ref{fig:4}). AAs outside the high shear region show only small epistasis, since mutations in the rigid domains hardly change the elastic response.
The epistasis landscapes (\ref{fig:4}A-C) are mostly positive since the mutations in the channel interact antagonistically \cite{Desai2007}: after a strongly deleterious mutation, a second mutation has a smaller effect. 

Definition [\ref{eq:epistasis}] is a direct link between epistasis and protein mechanics: The nonlinearity (`curvature')  of the Green function measures the deviation of the mechanical response from additivity of the combined effect of isolated mutations at $i$ and $j$,  $\DDGij \equiv \DGij-\DGi-\DGj$. 
The epistasis $\epij$ is simply the inner product value of this nonlinearity with the pinch and the response,
\begin{equation}
\label{eq:curvature}
 \epij = \vbT \, \DDGij \,  \fb~. 
\end{equation}
Relation [\ref{eq:curvature}] shows how epistasis originates from mechanical forces among mutated AAs.

In the gene, epistatic interactions are manifested in codon correlations \cite{Hopf2017,Poelwijk2017} shown in Fig.~\ref{fig:4}D, which depicts two-codon correlations $Q_{ij}$ from the alignment of ${\sim} 10^6$ functional genes $\cbs$ [\ref{eq:correlation}].
We find a tight correspondence between the mean epistasis $\EE_{ij}=\langle \epij \rangle$ and the codon correlations $Q_{ij}$. Both patterns exhibit strong correlations in the channel region with a period equal to channel's length, 10 AAs.
The similarity in the patterns of $Q_{ij}$ and $\EE_{ij}$ indicates that a major contribution to the long-range correlations observed among aligned protein sequences stems from the mechanical interactions propagating through the AA network.

\subsection*{Epistasis as a sum over scattering paths}
One can classify epistasis according to the interaction range. Neighboring AAs exhibit contact epistasis \cite{Goebel1994,Marks2011,Marks2012}, because two adjacent perturbations,  $\DHi$  and $\DHj$,  interact nonlinearly  via the ${\rm AND}$ gate of the interaction table (Fig.~\ref{fig:1}C), 
$\DDHij \equiv \DHij - \DHi  - \DHj \neq 0$ (where $\DHij$ is the perturbation by both mutations). The leading term in the Dyson series [\ref{eq:Dyson}] of  $\DDGij$ is a single scattering from an effective perturbation with an energy  $\DDHij$, which yields the epistasis 
\begin{equation*}
\epij = -\vbT \left[ \GG\, \DDHij  \GG  \right] \fb+\dotsb ~.
\end{equation*}
Long-range epistasis among non-adjacent, non-interacting perturbations  ($\DDHij=0$) is  observed along the  channel (Fig.~\ref{fig:4}). 
In this case,  [\ref{eq:Dyson}] expresses the nonlinearity  $\DDGij$ as a sum over multiple scattering paths which include both $i$ and $j$ (Fig. \ref{fig:2}C), 
 \begin{equation}
\epij = \vbT \left[
\GG\, \DHi \GG\,\DHj \GG +
\GG\, \DHj \GG\,\DHi \GG  \right] \fb 
 -\cdots~.
 \label{eq:multiple}
\end{equation}
The perturbation expansion directly links long-range epistasis to shear deformation:  Near the transition, the Green function is dominated by the soft mode, $\GG \simeq \ubs\ubsT/\lams$,
with fitness $F$ given by [\ref{eq:divergence}]. From [\ref{eq:Dyson}] and [\ref{eq:curvature}] we find a simple expression for the mechanical epistasis as a function of the shear,
\begin{equation}
\epij \simeq F \cdot \left[\frac{h_i}{1+h_i}+\frac{h_j}{1+h_j}-\frac{h_i+h_j}{1+h_i+h_j} \right]~.
\label{eq:antagonistic}
\end{equation}
The factor $h_i  \equiv \ubsT\DHi\ubs/\lams$ in [\ref{eq:antagonistic}] is the ratio of the change in the shear energy due to mutation at $i$ (the expectation value of $\DHi$) and the energy $\lams$ of the soft mode, and similarly for $h_j$. Thus, $h_i$ and $h_j $ are significant only in and around the shear band, where the bonds varied by the perturbations are deformed by the soft mode.  
When both sites are outside the channel, $h_i,h_j \ll 1$, the epistasis [\ref{eq:antagonistic}] is small, $\epij \simeq 2 h_i h_j F$. It remains negligible even if one of the mutations, $i$, is in the channel, $h_j \ll 1 \ll h_i$, and $\epij \simeq h_j F$. 
Epistasis can only be long-ranged along the channel when both mutations are significant, $h_i, h_j \gg 1$, and $\epij \simeq F \left[1 - h_i^{-1}-h_j^{-1} +(h_i+h_j)^{-1} \right] \simeq F \left[1 - 1/\min{(h_i,h_j)}\right]$.
We conclude that epistasis is maximal when both sites are at the start or end of the channel, as illustrated in Fig.~\ref{fig:4}. 
The nonlinearity of the fitness function gives rise to antagonistic epistasis.

\begin{figure*}[t]
\centering
\includegraphics[width=0.96\textwidth]{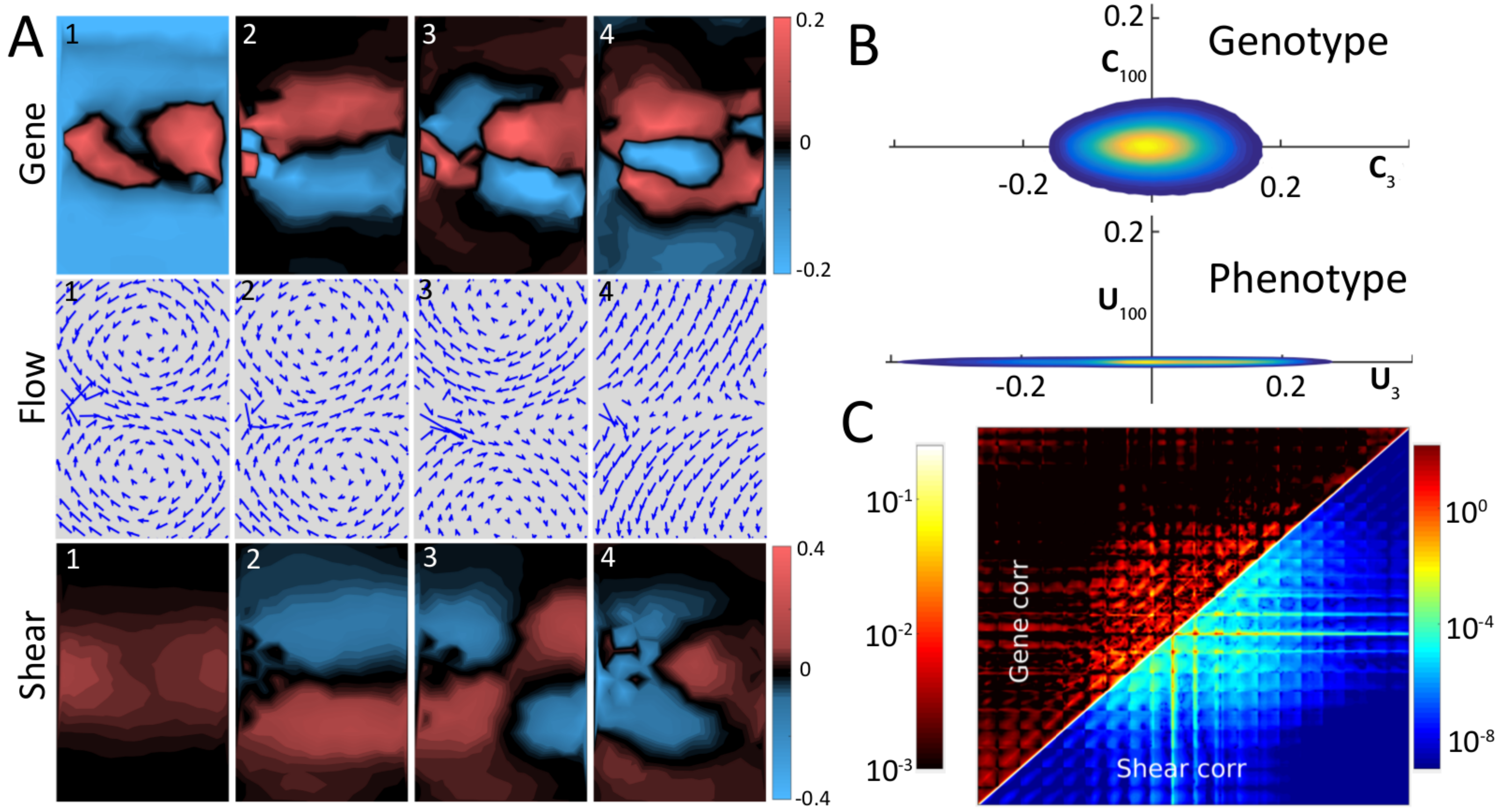}
\caption{\textbf{From gene to mechanical function: spectra and dimensions.} 
(A) The first four SVD eigenvectors (see text), of the gene $\Ck$ (top row), the displacement flow field $\Uk$ (middle) and the shear $\Sk$ (bottom).  
(B) Cross sections through the set of solutions in the genotype space (top) and the phenotype space (bottom). Density of solutions is color coded. The genotype cross section is the plane defined by the eigenvectors $\CC_3{-}\CC_{100}$, and in the phenotype space by the eigenvectors  $\UU_3{-}\UU_{100}$ (see text). 
The dimensional reduction is manifested by the discoid geometry of the phenotype cloud as compared to the spheroid shape of the genotype cloud.
(C)  Genetic correlations, $Q_{ij}$ show similarity to correlations in the shear field, $\sbs$ (color coded in log scale).}\label{fig:5}
\end{figure*}

\subsection*{Geometry of fitness landscape and gene-to-function map}
With our mechanical evolution algorithm we can swiftly explore the fitness landscape to examine its geometry. 
The genotype space is a $200$-dimensional hypercube, whose vertices are all possible genes $\cb$.  The phenotypes reside in a $400$-dimensional space of all possible mechanical responses $\ub$. And the Green function provides the genotype-to-phenotype map [\ref{eq:Green}].
A functional protein is encoded by a gene $\cbs$ whose fitness exceeds a large threshold, $F(\cbs) \ge F_{\rm m} \simeq 5$, 
and the functional phenotype is dominated by the emergent zero-energy mode, $\ub(\cbs) \simeq \ubs$ (Fig.~\ref{fig:3}A).  
We also characterize the phenotype by the shear field $\sbs$ (Methods). 

The singular value decomposition (SVD) of the $10^6$ solutions returns a set of eigenvectors, whose ordered eigenvalues show their significance in capturing  the data (Methods).  
The SVD spectra reveal strong correspondence between the genotype $\cbs$  and the phenotype, $\ub$ and $\sbs$ (Fig.~\ref{fig:5}). In all three data sets, the largest eigenvalues are discrete and stand out from the bulk continuous spectrum.  These are the collective degrees-of-freedom, which show loci in the gene and positions in the ``flow" (\ie displacement) and shear fields that tend to vary together. 

We examine the correspondence among three sets of eigenvectors, $\{\Uk\}$ of the flow, $\{\Ck\}$ of the gene, and $\{\Sk\}$ of the shear.
The first eigenvector of the flow, $\UU_1$, is the hinge motion caused by the pinch, with two eddies rotating in opposite directions (Fig.~\ref{fig:5}A). The next two modes, shear, $\UU_2$, and breathing, $\UU_3$, also occur in real proteins such as glucokinase (Fig.~\ref{fig:1}B). 
The first eigenvectors of the shear  $\SS_1$ and of the gene sequence $\CC_1$ show that the high shear region is mirrored as a $\aP$-rich region, where a mechanical signal may cause local rearrangement of the AAs  by deforming weak bonds. In the rest of the protein, the $\aH$-rich regions move as rigid bodies with insignificant shear.
The higher gene eigenvectors, $\Ck$ ($k >1$), capture patterns of correlated genetic variations. The striking similarity between the sequence correlation patterns $\Ck$  and  the shear eigenvectors $\Sk$ shows a tight genotype-to-phenotype map, as is further demonstrated in the likeness of the correlation matrices of the AA and shear flow (Fig.~\ref{fig:5}C).

In the phenotype space, we represent  the displacement field $\ub$  in the SVD basis,  $\{\Uk\}$ (Fig.~\ref{fig:5}B). 
Since ${\sim}90\%$ of the data is explained by the first ${\sim}15$ $\Uk$, we can compress the displacement field without much loss into the first 15 coordinates.  
This implies that the set of solutions is a $15$-dimensional discoid which is flat in most directions. 
In contrast, representation of the genes $\cbs$ in the SVD frame-of-reference (with the $\{\Ck\}$ basis) reveals that in genotype space the solution set is an incompressible $200$-dimensional spheroid (Fig.~\ref{fig:5}B). 
The dramatic dimensional reduction in mapping genotypes to phenotypes stems from the different constraints that shape them \cite{Tlusty2017,Savir2010,Savir2013,Kaneko2015,Friedlander2015,Furusawa2017,Tlusty2010}. 

\section*{Discussion}
 
Theories of protein need to combine the many-body physics of the amino acid matter with the evolution of genetic information, which together give rise to protein function. 
We introduced a simplified theory of protein, where the mapping between genotype and phenotype takes an explicit form in terms of mechanical propagators (Green functions), which can be efficiently calculated. 
As a functional phenotype we take cooperative motion and force transmission through the protein [\ref{eq:fitness}]. This allows us to map genetic mutations to mechanical perturbations which scatter the force field and deflect its propagation [\ref{eq:deltaF},\ref{eq:Dyson}] (Fig. \ref{fig:2}). 
The evolutionary process amounts to solving the inverse scattering problem: Given prescribed functional modes, one looks for network configurations, which yield this low end of the dynamical spectrum. 
Epistasis, the interaction among loci in the gene, corresponds to a sum over all multiple scattering trajectories, or equivalently the nonlinearity of the Green function [\ref{eq:epistasis},\ref{eq:curvature}]. We find that long-range epistasis signals the emergence of a collective functional mode in the protein. 
The results of the present theory, in particular the expressions for epistasis, follow from the basic geometry of the AA network and the localized mutations, and are therefore applicable to general tasks and fitness functions with multiple forces and responses.

\vfill\null

\matmethods{
\subsection*{The mechanical model of protein}
We model the protein as an elastic network made of harmonic springs \cite{Born1954,Alexander1998,Chennubhotla2005,Tirion1996}. The connectivity  of the network is described by a hexagonal lattice whose vertices are AAs and whose edges correspond to bonds. There are $n_a=10 \times 20=200$ AAs, indexed by Roman letters and $n_b$ bonds, indexed by Greek letters. 
We use the HP model \cite{Dill1985} with two AA species, hydrophobic ($a_i = \aH$) and polar ($a_i=\aP$). The AA chain is encoded in a gene $\cb$,  where $c_i=1$  encodes $\aH$ and   $c_i=0$ encodes $\aP$, $i=1,\dots, n_a$. The degree $z_i$ of each AA is the number of AAs to which it is connected by bonds. In our model, most AA have the maximal degree which is $z=12$, while AAs at the boundary have fewer neighbors, $z<12$, see Fig.~\ref{fig:1}C.   
The connectivity of the graph is recorded by the adjacency matrix $\Ab$, where $\Ab_{ij}=1$ if there is a bond from $j$ to $i$, and $\Ab_{ij}=0$ otherwise. The gradient operator $\nabla$ relates the spaces of bonds and vertices (and is therefore of size $n_b {\times} n_a$): if vertices $i$ and $j$ are connected by a bond $\alpha$, then $\nabla_{ \alpha i} =
+1$ and $\nabla_{\alpha j } =- 1$.   As in the continuum case, the Laplace operator $\Delta$ is the product $\Delta = \nabla^\T \nabla$. The non-diagonal elements $\Delta_{ij}$ are $-1$ if $i$ and $j$ are connected and $0$ otherwise. The diagonal part of $\Delta$ is the degree  $\Delta_{ii}= z_i$. Hence, we can write the Laplacian as $\Delta = \Zb - \Ab$, where $\Zb$ is the diagonal matrix of the degrees $z_i$. 

We embed the graph in Euclidean space  $\mathbb{E}^d$ ($d=2$), by assigning positions $\rb_i \in \mathbb{E}^d$ to each AA .  We concatenate all positions in a vector $\rb$ of length $n_a \cdot d 
\equiv n_d$. Finally, to each bond we assign a spring with constant $k_{\alpha}$, which we keep in a diagonal $n_b {\times} n_b$ matrix $\Kb$. The strength of the spring is determined by the ${\rm AND}$ rule of the HP model's interaction table (Fig.~\ref{fig:1}C), $k_\alpha = k_w + (k_s-k_w) c_i c_j$, where $c_i$ and $c_j$ are the are the codons of the AA connected by bond $\alpha$. This implies that a strong $\aH{-}\aH$ bond has  $k_\alpha=k_s$, whereas the weak bonds $\aH{-}\aP$ and $\aP{-}\aH$ have $k_\alpha=k_w$. We usually take $k_s =1$ and $k_w=0.01$. This determines a spring network. We also assume that the initial configuration is such that all springs are at their equilibrium length, disregarding the possibility of  `internal stresses' \cite{Alexander1998}, so that the initial elastic energy is $\mathcal{E}_0=0$.  

We define the `embedded' gradient operator  $\Db$ (of size $n_b {\times} n_d)$ which is obtained by taking the graph gradient $\nabla$, and multiplying  each non-zero element $(\pm1)$ by the corresponding direction vector $\nij= \left(\rb_i - \rb_j\right)/\left\vert\rb_i -
\rb_j\right\vert$. Thus, $\Db$ is a tensor ($\Db= \Delta_{\alpha
  i}\nij$), which we store as a matrix ($\alpha$ is the bond connecting vertices $i$ and $j$).  In each row vector of $\Db$, which we denote as $\mba \equiv\ \Db_{\alpha, :}$, there are only $2d$ non-zero elements. 
To calculate the elastic response of the network, we deform it  by applying a force field $\fb$, which leads to the displacement of each vertex by $\ub_i$ to a new position $\rb_i+\ub_i$ (see \eg \cite{Alexander1998}). For small displacements, the linear response of the network is given by Hooke's law,  $\fb = \HH \ub$.  
The elastic energy is $\mathcal{E}=  \ub^\T \HH \ub/2$, and the Hamiltonian, $\HH=\Db^\T \Kb\Db$, is the Hessian of the elastic energy $\mathcal{E}$, $\HH_{ij}=\delta^2\mathcal{E} /(\delta \ub_i \delta \ub_j)$. 
By rescaling, $\Db \to \Kb^{1/2} \Db$, which amounts to scaling all distances by $1/\sqrt{k_{\alpha}}$, we obtain $\HH = \Db^\T \Db$. It follows that the Hamiltonian  is a function of the gene $\HH(\cb)$, which has the structure of the Laplacian $\Delta$ multiplied by the tensor product of the direction vectors. Each $d {\times} d$ block $\HH_{ij}$ ($i \neq j$) is a function of the codons $c_i$ and $c_j$,
\begin{equation}
\begin{aligned}
\HH_{ij} (c_i,c_j) & = \Delta_{ij} \nij \nijT   \\
&= -\Ab_{ij} \left[k_w+ \left(k_s-k_w\right) c_i c_j\right] \nij \nijT~.
\label{eq:Hamiltonian}
\end{aligned}
\end{equation}
The diagonal blocks complete the row and column sums to zero, $\HH_{ii}=-\sum_{j \ne i}{\HH_{ij}}$. 

\subsection*{The inverse problem: Green's function and its spectrum}
 The Green function $\GG$ is defined by the inverse relation to Hooke's law, $\ub = \GG \fb$ [\ref{eq:Green}]. 
In the physics literature, the  term `Green function' often refers to the two-point response functions $\GG_{ij}$  which quantify the response at $i$ to a point force at $j$ or vice versa (these are $n_a{\times} n_a$ numbers for all direction combinations). 
If $\HH$ were invertible (non-singular),$\GG$ would have been just $\GG=\HH^{-1}$. However, $\HH$ is always singular owing to the zero-energy (Galilean) modes of translation and rotation. Therefore, one needs to define $\GG$ as the Moore-Penrose pseudo-inverse \cite{Penrose1955,Ben-Israel2003}, $  
\GG=\HH^\pseudo$, on the complement of the space of Galilean transformations.
The pseudo-inverse can be understood in terms of the spectrum of $\HH$. There are at least $n_0= d(d+1)/2$ zero modes: $d$ translation modes and $d(d-1)/2$ rotation modes. These modes are irrelevant and will be projected out of the calculation (Note that these modes do not come from missing connectivity of the graph $\Delta$ itself but from its embedding in $\mathbb{E}^d$). $ \HH$ is singular but is still diagonalizable (since it has a basis of dimension $n_d$), and can be written as the spectral decomposition, $ \HH = \sum_{k=1}^{n_d}{\lambda_k}{\ub_{k}\ub_{k}^\T}$, where $\{\lambda_k\}$ is the set of eigenvalues and $\{\ub_{k}\}$ are the corresponding eigenvectors (note that $k$ denotes the index of the eigenvalue, while $i$ and $j$ denote AA positions).
For a non-singular matrix one may calculate the inverse simply as $\HH^{-1} = \sum_{k=1}^{n_d}{\lambda_k^{-1}}{\ub_{k}\ub_{k}^\T}$. Since $\HH$ is singular we leave out the zero modes and get the pseudo-inverse $\HH^{\pseudo}$,  $  \GG = \HH^{\pseudo}=\sum_{k=n_0+1}^{n_d} {\lambda_k^{-1} \ub_{k}\ub_{k}^\T}$.
It is easy to verify that if $\ub$ is orthogonal to the zero modes then $\ub = \GG \HH\ub$. The pseudo-inverse obeys the four requirements \cite{Penrose1955}:  (i)  $\HH \GG \HH = \HH$, (ii) $\GG \HH \GG = \GG$, (iii) $(\HH \GG)^\T = \HH \GG$, (iv) $(\GG \HH)^\T = \GG \HH$. 
In practice, as the  projection commutes with the mutations, the pseudo-inverse has most  virtues of a proper inverse.
The reader might prefer to link $\GG$ and $\HH$ through the heat kernel,  $\KK(t) = \sum_k {e^{\lambda_k t} \ub_{k} \ub_{k}^{\T}}$. Then,  $\GG = \int_0^\infty{\d t~\KK(t)}$ and $\HH=\frac{\d}{\d t}\,{\KK}|_{t=0}$~.

\subsection*{Pinching the Network}
A pinch is given as a localized force applied at the boundary of the `protein'. We usually apply the force on a pair of neighboring boundary vertices, $\bi$ and $\bj$. It appears reasonable to apply a force dipole, \ie   two opposing forces, $\fb_\bj=-\fb_\bi$, since a net force  will move the center of mass. This `pinch' is therefore specified by the force vector $\fb$ (of size $n_d$) whose only $2d$ non-zero entries are $\fb_\bj=-\fb_\bi$.  
Hence, it has the same structure as a bond vector $\mba$ of a `pseudo-bond' connecting $\bi$ and $\bj$ A normal `pinch' $\fb$ has a force dipole directed along the  $\rb_\bi-\rb_\bj$  line (the $\nb_{\bi\bj}$ direction). Such a pinch is expected to induce a hinge motion. A shear `pinch' will be in a perpendicular direction $\perp\nb_{\bi\bj}$ , and is expected to induce a shear motion.   

In the protein scenario, we try to tune the spring network to exhibit a low-energy mode in which the protein is divided into two sub-domains moving like rigid bodies.  This large-scale mode can be detected by examining the relative motion of two neighboring  vertices, $\ui$ and $\uj$ at another location at the boundary (usually at the opposite side). Such a desired response at the other side of the protein is specified by a response vector $\vb$, whose only non-zero entries correspond to the directions of the response at  $\ui$ and $\uj$ . Again, we usually consider a `dipole' response $\vb_\uj=-\vb_\ui$.

\subsection*{Evolution and Mutation}
The quality of the response, \ie    the biological fitness, is 
specified by how well the response follows the prescribed one $\vb$. In
the context of our model, we chose the (scalar) observable $F$ as $F=\vbT\ub =\vb_{\ui}\ub_{\ui}+\vb_{\uj}\ub_{\uj} =\vbT \GG \fb$ [\ref{eq:fitness}].
In an evolution simulation one would exchange AAs between $\aH$ and $\aP$ while demanding that the fitness change $\delta F$ is positive or non-negative. By this, we mean $\delta F >0$ is thanks to a beneficial mutation, whereas $\delta F=0$ corresponds to a neutral one. Deleterious mutations $\delta F<0$ are generally rejected. A version which accepts mildly deleterious mutations (a finite-temperature Metropolis algorithm) gave similar results.
 We may impose a stricter minimum condition, $\DF \ge \varepsilon\ F$ with a small positive $\varepsilon$, say $1\%$. 
 An alternative, stricter criterion would be the demand that each of the terms in $F$, $\vb_{\ui}\ub_{\ui}$ and $\vb_{\uj}\ub_{\uj} $, increases separately. 
The evolution is stopped when $F \ge F_{\rm m} \sim 5$, which signals the formation of a shear band.  When simulations ensue beyond $F_{\rm m}\sim 5$, the band slightly widens and the fitness slows down and converges at a maximal value, typically $F_{\rm max} \sim 8$.

\subsection*{Evolving the Green function using Dyson's and Woodbury's formulas}
Dyson's formula follows from the identity $\DH \equiv \Hs - \HH = \Gs^\pseudo - \GG^\pseudo$, which is multiplied by $\GG$ on the left and $\Gs$ on the right to yield [\ref{eq:Dyson}].
The formula remains valid for the pseudo-inverses in the non-singular sub-space. 
One can calculate the change in fitness by evaluating the effect of a mutation on the Green function, $\Gs = \GG +\DG$, and then examining the change,  $\DF = \vbT \DG \fb$ [\ref{eq:deltaF}]. 
Using [\ref{eq:Dyson}] to calculate the mutated Green function $\Gs$  is an impractical method  as it amounts to inverting at each step a large $n_d {\times} n_d$ matrix. 
However, the mutation of an AA at $i$ has a localized effect. It may change only up to $z=12$ bonds among the bonds $\alpha(i)$ with the neighboring AAs. Thanks to the localized nature of the mutation,  the corresponding defect Hamiltonian $\DHi$  is therefore of a small rank, $r \le z =12$, equal to the number of switched bonds (the average $r$ is about 9.3).  
$\DHi$ can be decomposed into a product $\DHi= \MM\Bb\MM^\T$. The diagonal $r {\times} r$ matrix $\Bb$ records whether a bond $\alpha(i)$ is switched from weak to strong ($\Bb_{\alpha \alpha}=k_s-k_w=+0.99$) or vice versa ($\Bb_{\alpha \alpha}=-0.99$), and $\MM$ is a $n_d {\times} r$ matrix whose $r$ columns are the  bond vectors $\mba$ for the switched bonds $\alpha(i)$ . 
This allows one to calculate changes in the Green function more efficiently using the Woodbury formula \cite{Woodbury1950,Deng2011},
 \begin{equation} \label{eq:Woodbury}
\DG= -\GG\MM \left(\Bb^{-1} + \MM^{\T}\GG\MM \right)^{\pseudo}\MM^\T\GG~.
\end{equation}

The two expressions for the mutation impact $\DG$, [\ref{eq:Dyson}] and [\ref{eq:Woodbury}], are equivalent and one may get the scattering series of [\ref{eq:Dyson}] by a series  expansion of the pseudo-inverse in [\ref{eq:Woodbury}]. The practical advantage of [\ref{eq:Woodbury}] is that the only (pseudo-) inversion it requires is of a low-rank tensor (the term in brackets). This accelerates our simulations by a factor of $(n_a/ r)^3 \simeq 10^4$.

\subsection*{Pathologies and broken networks}
A network broken into disjoint components exhibits floppy modes owing to the low energies of the relative motion of the components with respect to each other. The evolutionary search might end up in such non-functional unintended modes. The common pathologies we observed are: (i) isolated nodes at the boundary that become weakly connected via  $\aH {\to }\aP$ mutations, (ii) `sideways' channels that terminate outside the target region (which typically includes around $8{-}10$ sites), and (iii) channels that start and end at the target region without connecting to the binding site.  All these are some easy-to-understand floppy modes, which can vibrate independently of the location of the pinch and cause the response to diverge (${>}F_{\rm m}$) without producing a functional mode. We avoid such pathologies by applying the pinch force to the protein network symmetrically: pinch the binding site on face L and look at responses on face R and vice versa. Thereby we not only look for the transmission of the pinch from the left to right but also from right to left.  The basic algorithm  is modified to accept a mutation only if it does not weaken the two-way responses and enables hinge motion of the protein. This prevents the vibrations from being localized at isolated sites or unwanted channels.

\subsection*{Dimension and Singular Value Decomposition (SVD)}
To examine the geometry of the fitness landscape and the genotype-to-phenotype map, we looked at the correlation among numerous solutions, typically $N_{\rm sol} \sim 10^6$.
Each solution is characterized by three vectors: (i) the gene of the functional protein, $\cbs$, (a vector of length $n_a=200$ codons) (ii) the flow field (displacement),  $\ub(\cbs)=\GG(\cbs) \fb$, (a vector of length $n_d =400$ of $x$ and $y$ velocity components),  (iii)  the shear field $\sbs$ (a vector of length  $n_a=200$). We compute the shear as the symmetrized derivative of the displacement field using the method of \cite{Mitchell2016}.  The values of the $\sbs$ field is the sum of squares of the traceless part of the strain tensor (Frobenius norm).  
These three types of vectors are stored along the rows of three matrices $W_C$, $W_U$ and $W_S$. 
We calculate the eigenvectors of these matrices, $\Ck$, $\Uk$ and $\Sk$,  via singular value decomposition (SVD) (as in \cite{Tlusty2017}). The corresponding SVD eigenvalues are the square roots of the eigenvalues of the covariance matrix  $W^\T W$, while the eigenvectors are the same. In typical spectra, most eigenvalues reside in a continuum bulk region that resembles the spectra of random matrices. A few larger outliers, typically around a dozen or so, carry the non-random correlation information. 

\begin{figure}[htb]
\centering
\includegraphics[width=0.5\textwidth]{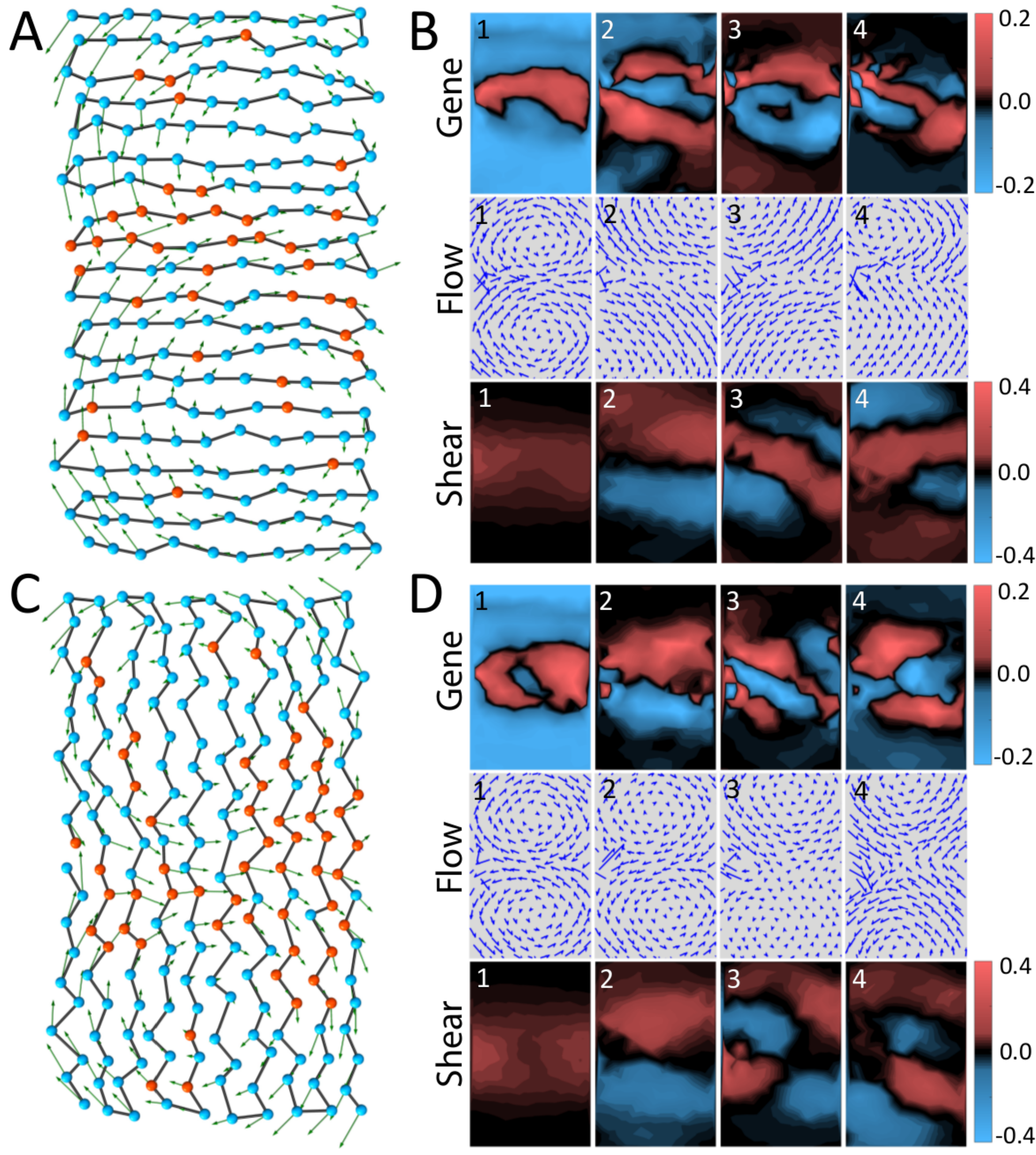}
\caption{\textbf{The effect of the backbone on evolution of mechanical function.} 
The backbone induces long-range mechanical correlations which influence protein evolution. 
We examine two configurations:  parallel (A-B) and perpendicular to the channel (C-D). 
Parallel: (A) The backbone directs the formation of a narrow channel along the fold (compared to Fig.~\ref{fig:5}A).
(B) First four SVD eigenvectors of the gene $\Ck$ (top row), the flow $\Uk$ (middle) and the shear $\Sk$ (bottom).
Perpendicular: (C) The formation of the channel is `dispersed' by the backbone. 
(D). First four SVD eigenvectors of $\Ck$ (top row),  $\Uk$ (middle) and shear $\Sk$ (bottom).}
\label{fig:6}
\end{figure}

\subsection*{The protein backbone}
A question may arise as to what extent the protein's backbone might affect the results described so far.
Proteins are polypeptides, linear heteropolymers of AAs, linked by covalent peptide bonds, which form the protein backbone. 
The peptide bonds are much stronger than the non-covalent interactions among the AAs and do not change when the protein mutates.  
We therefore augmented our model with a `backbone': a linear path of conserved strong bonds that passes once through all AAs. 
We focused on two extreme cases, a serpentine backbone either parallel to the shear band or perpendicular to it (Fig.~\ref{fig:6}). 

The presence of the backbone does not interfere with the emergence of a low-energy mode of the protein whose flow pattern (\ie displacement field) is similar to the backbone-less case with two eddies moving in a hinge like fashion.  
In the parallel configuration, the backbone constrains the channel formation to progress along the fold (Fig.~\ref{fig:6}A).  
The resulting channel is narrower than in the model without backbone (Figs. \ref{fig:1}D, \ref{fig:5}).
In the perpendicular configuration, the evolutionary progression of the channel is much less oriented (Fig.~\ref{fig:6}C). While the flow patterns are similar, closer inspection shows noticeable differences, as can be seen in the flow eigenvectors $\Uk$ (Fig.~\ref{fig:6}B,D). The shear eigenvectors $\Sk$ represent  the derivative of the flow, and therefore highlight more distinctly these differences. 

As for the correspondence between gene eigenvectors $\Ck$ and shear eigenvectors $\Sk$, the backbone affects the shape of the channel in concert with the sequence correlations around it. 
Transmission of mechanical signals appears to be easier along the orientation of the fold (parallel configuration, Fig.~\ref{fig:6}A). Transmission across the fold (perpendicular configuration) necessitates significant deformation of the backbone and leads to `dispersion' of the signal at the output (Fig.~\ref{fig:6}C). 
We propose that the shear band will be roughly oriented with the direction of the fold, but this requires further analysis of structural data. Overall, we conclude from our examination that the backbone adds certain features to patterns of the field and sequence correlation, without changing the basic results of our model. 
The presence of the backbone might constrain the evolutionary search, but this has no significant effect on the fast convergence of the search and on the long-range correlations among solutions.

}
\showmatmethods{} 
\acknow{We thank Jacques~Rougemont for calculations of shear in glucokinase (Fig.~\ref{fig:1}B) and for helpful discussions.
We thank Stanislas~Leibler, Michael R.~Mitchell, Elisha~Moses, Giovanni~Zocchi and Olivier~Rivoire  for helpful discussions and encouragement.
We thank Alex~Petroff, Steve~Granick, Le Yan and Matthieu Wyart  for valuable comments on the manuscript. 
JPE is supported by an ERC advanced grant `Bridges', and TT by the Institute for Basic Science IBS-R020 and the Simons Center for Systems Biology of the Institute for Advanced Study, Princeton.
}
\showacknow{} 

\bibliography{green}

\end{document}